\documentclass[twocolumn,showpacs,preprintnumbers,amsmath,amssymb,aps,floatfix]{revtex4-1}

\usepackage{amssymb,amsmath,graphicx,bm,textcomp,color}

\newcommand{\jc}{\color{black}}

\begin{document}

\title{Pressure dependent friction on granular slopes close to avalanche}

\author{J\'er\^ome Crassous$^{1}$\email{jerome.crassous@univ-rennes1.fr}}
\author{Antoine Humeau$^{2,3}$}
\author{Samuel Boury$^{1,4}$}
\author{J\'er\^ome Casas$^{2,5}$\email{jerome.casas@univ-tours.fr}}

\affiliation{$^{1}$Universit\'e Rennes 1, Institut de Physique de Rennes (UMR UR1-CNRS 6251),Campus de Beaulieu, F-35042 Rennes, France}
\affiliation{$^{2}$Institut de Recherche sur la Biologie de l'Insecte, UMR 7261 CNRS - Universit\'e Fran\c{c}ois-Rabelais, 37200 Tours, France}
\affiliation{$^{3}$Office National de la Chasse et de la Faune Sauvage, Direction de la Recherche et de l\textquotesingle Expertise, Unit\'e Sanitaire de la Faune, F-78610 Auffargis, France}
\affiliation{$^{4}$\'Ecole Normale Sup\'erieure de Lyon, Lyon, France}
\affiliation{$^{5}$Institut Universitaire de France}


\begin{abstract} We investigate the sliding of objects on an inclined granular surface close to the avalanche threshold. Our experiments show that the stability is driven by the surface deformations. Heavy objects generate footprint-like deformations which stabilize the objects on the slopes. Light objects do not disturb the sandy surfaces and are also stable. For intermediate weights, the deformations of the surface generate a sliding of the objects. The solid friction coefficient does not follows the Amontons-Coulomb laws, but is found minimal for a characteristic pressure. Applications to the locomotion of devices and animals on sandy slopes as a function of their mass are proposed.
\end{abstract}

\maketitle

Sandy slopes are slippery in the sense that objects or animals may easily slide on it. A remarkable example of this tendency to slide down are prey trapped in antlion traps. The antlion larva dig into sand inverted cone at an angle close to the avalanche angle. Ants entering into such cones slide down towards their predators located at the bottom of the cone.~\cite{Griffiths1980a,Fertin2006}.
\begin{figure}[tbh]
\includegraphics[width=.55\columnwidth]{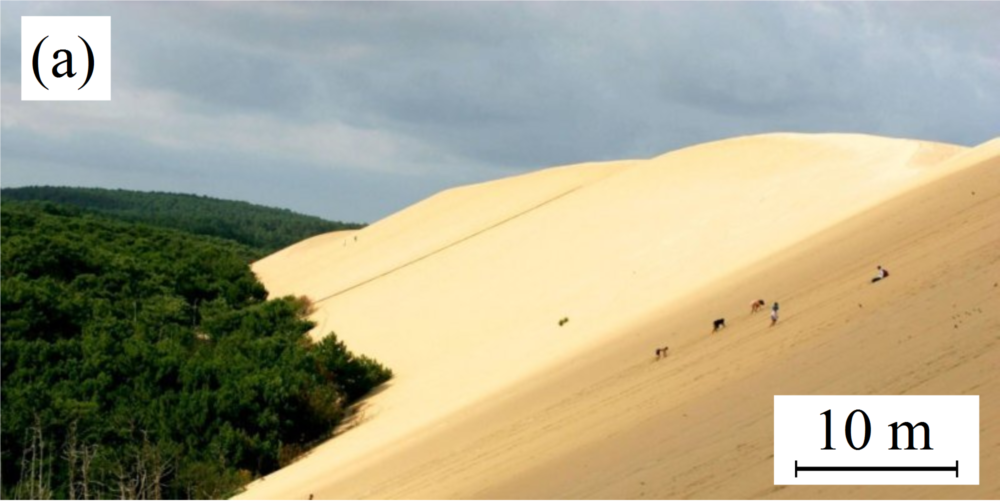}%
\includegraphics[width=.45\columnwidth]{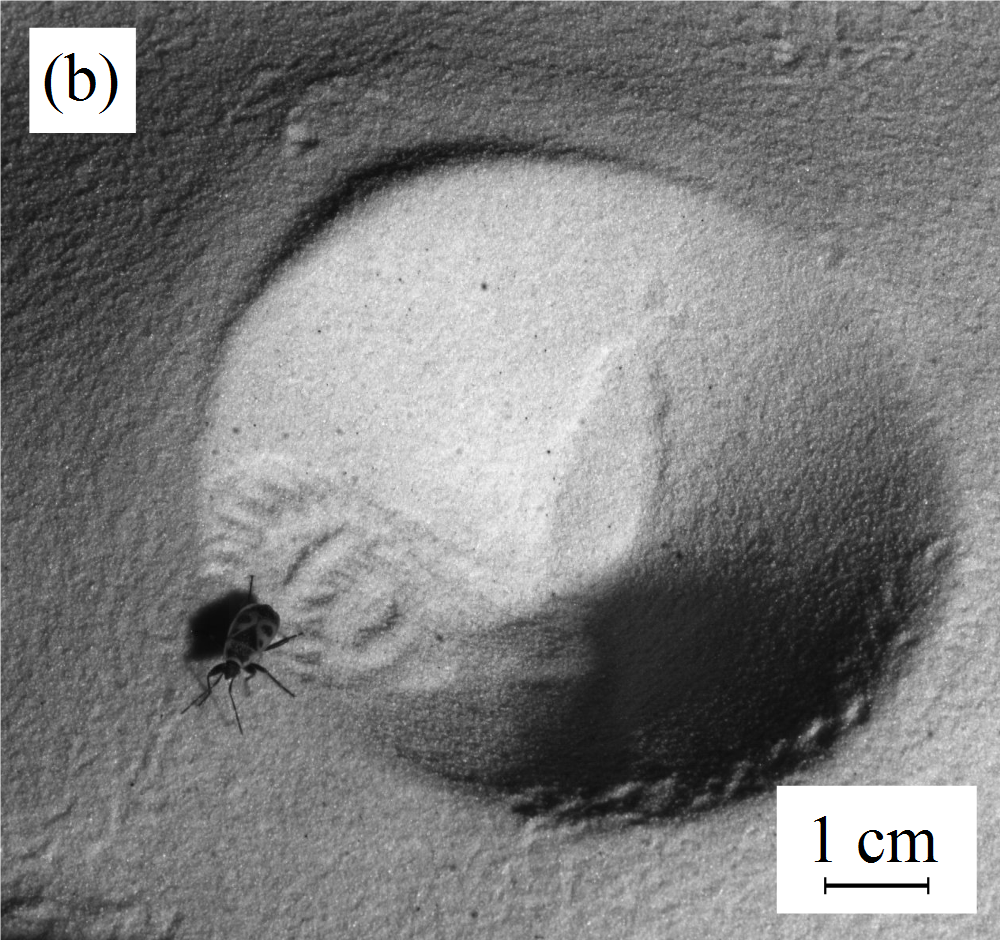}%
\caption{(a) Humans climbing the slip face of the "dune du Pilat" (France). Photo Franck Perrogon, with permission. (b) Firebug escaping from an antlion trap. On both pictures, the angle of the granular materials is close to the avalanche angle.}
\label{fig1}
\end{figure}
However, as shown in fig.1, climbing sandy slopes at angles close to the avalanche angle is possible for beings larger than ants, such as human or firebugs. It is indeed known that the ability of an insect to escape from an antlion trap depends on its weight~\cite{Humeau.etal2015}.

More generally, understanding the mechanism of solid friction over sandy materials is of importance for various fields that include animal locomotion~\cite{li2013,hosoi2015, marvi2014}, civil engineering~\cite{nedderman} and interaction between mechanical devices and soils~\cite{taberlet2007,taberlet2011,li2013,hosoi2015}. On a horizontal sandy surface, tangential forces are proportional to the normal forces~\cite{geminard1999,fall2014}, i.e. the Amontons's laws of friction~\cite{amontons1699} roughly describe the frictional forces. To our knowledge, the laws of friction over inclined granular surfaces are unknown. In particular there is no evidence that the frictional force exerted on objects or animals can be described by Amontons laws.

In this letter, we investigate experimentally the resistance to sliding on inclined granular surfaces as a function of object's weight. We found that the sliding ability increases close to the avalanche angle. Moreover, the sliding ability evolves strongly with the weights of the objects, and  a complex diagram of stability emerges from the experiments. The ability of objects to deform dynamically a granular surface appears to be a key parameter that control the stability of objects located on slopes. Finally we propose a link between this diagram of stability and the inability of some insects to escape from an antlion trap.

\begin{figure}[tbh]
\includegraphics[width=1.\columnwidth]{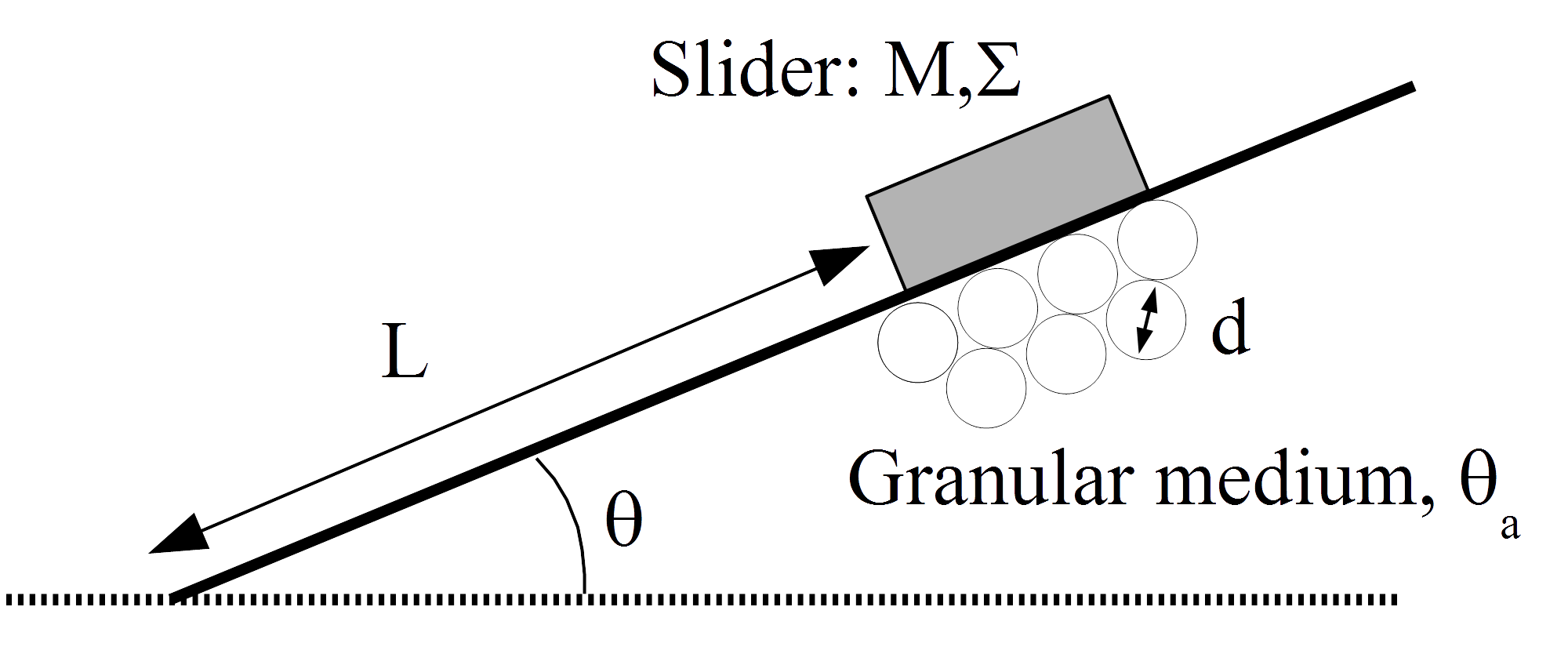}
\caption{A granular material (particles' size $d$, avalanche angle $\theta_a$) is inclined at an angle $\theta$. A slider of mass $M$ and bottom surface $\Sigma$ is gently deposited on the surface, and may slide over a maximal distance $L$.}
\label{fig2}
\end{figure}

The experimental setup is schematically drawn in fig.2. A granular material is prepared by pouring glass beads {\jc (density $\rho=2.50\times 10^3~kg.m^{-3}$)} of diameter $d=1.2~mm$ (mean avalanche angle $\theta_a=27~deg$), $d=2.2~mm$ ($\theta_a=29~deg$) and $d=6.0~mm$ ($\theta_a=29~deg$) into a box (length = $29~cm$, width = $20~cm$, height = $4~cm$), then the surface is leveled. The solid volume fraction is $\phi \simeq 0.60$. The box is then inclined at an angle $\theta$. We place objects (mass $M$, surface $\Sigma$) delicately on the surface, and objects can slide down, then eventually stop. A picture is then taken and the sliding distances are measured. The avalanche angle is then measured and new granular surface is prepared. For an object with a given mass and surface, the fact that an object slides or stops is not deterministic. Experiments are repeated $8$ or $16$ times depending on the physical parameters. The objects are small circular metallic pieces with a cardboard's surface. The sliding friction coefficient $\mu$ of objects on a layer of glued beads is measured as $\mu=\tan(22~deg)=0.40$.

The outcome of such experiments is the  probability $\varphi$ that the object slides over the box length. Dimensionless governing quantities are the reduced surface $\Sigma^*=\Sigma/d^2$ and the relative pressure $P^*=P/\rho g d$, with $g$ the gravity and $P=Mg/\Sigma$ the pressure exerted by the object on a horizontal surface.

\begin{figure}[!tbh]
\includegraphics[width=\columnwidth]{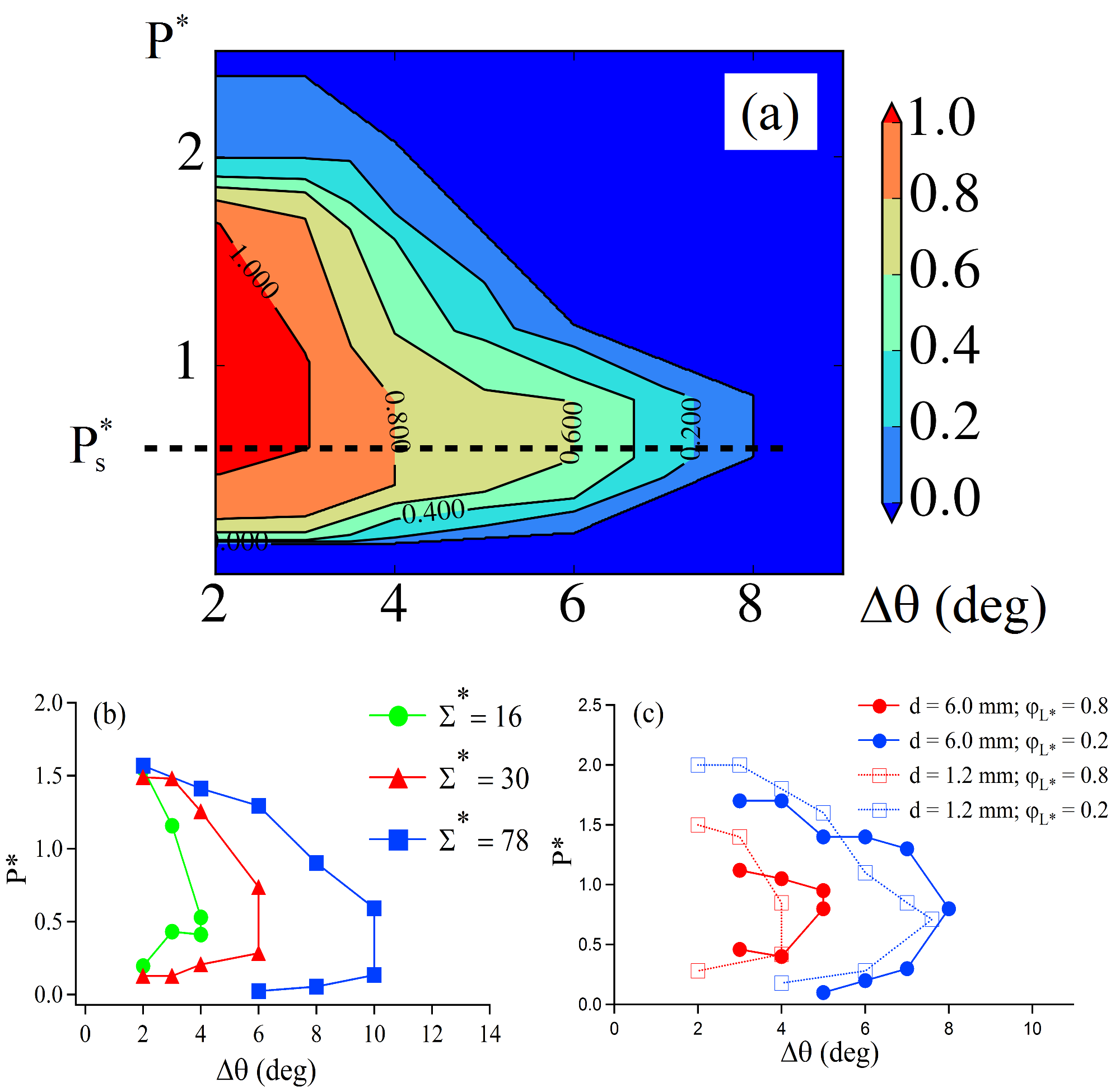}
\caption{(a) Curves of constant probability of sliding in the $(\Delta \theta, P^*)$ plane for $d=2.2~mm$, $\Sigma^*=30$. $P_s^*$ is the value of $P^*$ for which the sliding probability is maximum. (b) Sliding probabilities $\varphi_{L^*}=0.4$ in the $(\Delta \theta, P^*)$ plane for different reduced surfaces $\Sigma^*$ ($d=2.2~mm$). (c) Sliding probabilities for two diameters of beads ($\Sigma^*=30$).}
\label{fig3}
\end{figure}

We first discuss the variations of the stability obtained by changing $P^*$ and $\Delta \theta=\theta_a-\theta$ the angular distance to avalanche. The probability $\varphi$ is evaluated for every degree of $\Delta \theta$ between $2$ and $8~\deg$, and for various values of $P^*$. The curves of constant sliding probability are obtained by interpolation between those data (fig.3(a)). We first observe that sliding is restricted to a quite small area at the left of the $(\Delta \theta, P^*)$ plane. As expected, the decrease of $\varphi_{L^*}$ with $\Delta \theta$ reflects that the stability decreases as the granular surface inclination approaches the avalanche angle. For a given $\Delta \theta$, the stability depends on $P^*$.

 The stability diagram is robust with respect of physical parameters variations. We performed experiments for various $\Sigma^*$, and we report in fig.3(b) the iso-probability curves $\varphi=0.4$. The existence of a zone stability with the same shape occurs for any $\Sigma^*$. The pressure $P^*_s$ corresponding to the maximum of sliding appears to be roughly independent of $\Sigma^*$. Also, the instability region extends to larger values of $\Delta \theta$ when $\Sigma^*$ increases. {\jc  We also verified that the stability diagrams are similar when the sizes of the beads and of the slider are up-scaled by a factor five (see Figure 3(c))}.

Remarkably, the instability is limited to a finite range of pressure: objects exerting a small or large pressure do not slide. By contrast, objects exerting intermediate pressures slide. The sliding probability that we measure depends on the applied pressure $P^*$. If the friction coefficient of the slider follows the Amontons-Coulomb law, the constant probability should then be vertical lines, which is not the case here. The stability diagram does not seems related to the triggering of granular avalanches~\cite{daerr1999}. Indeed, for $\Delta \theta \ge 2~deg$, we never observed granular avalanches. Moreover, the maximum and minimum angles of stability of granular slopes are invariant with length scales, and thus independent of $P^*$.

\begin{figure}[!th]
\includegraphics[width=0.9\columnwidth]{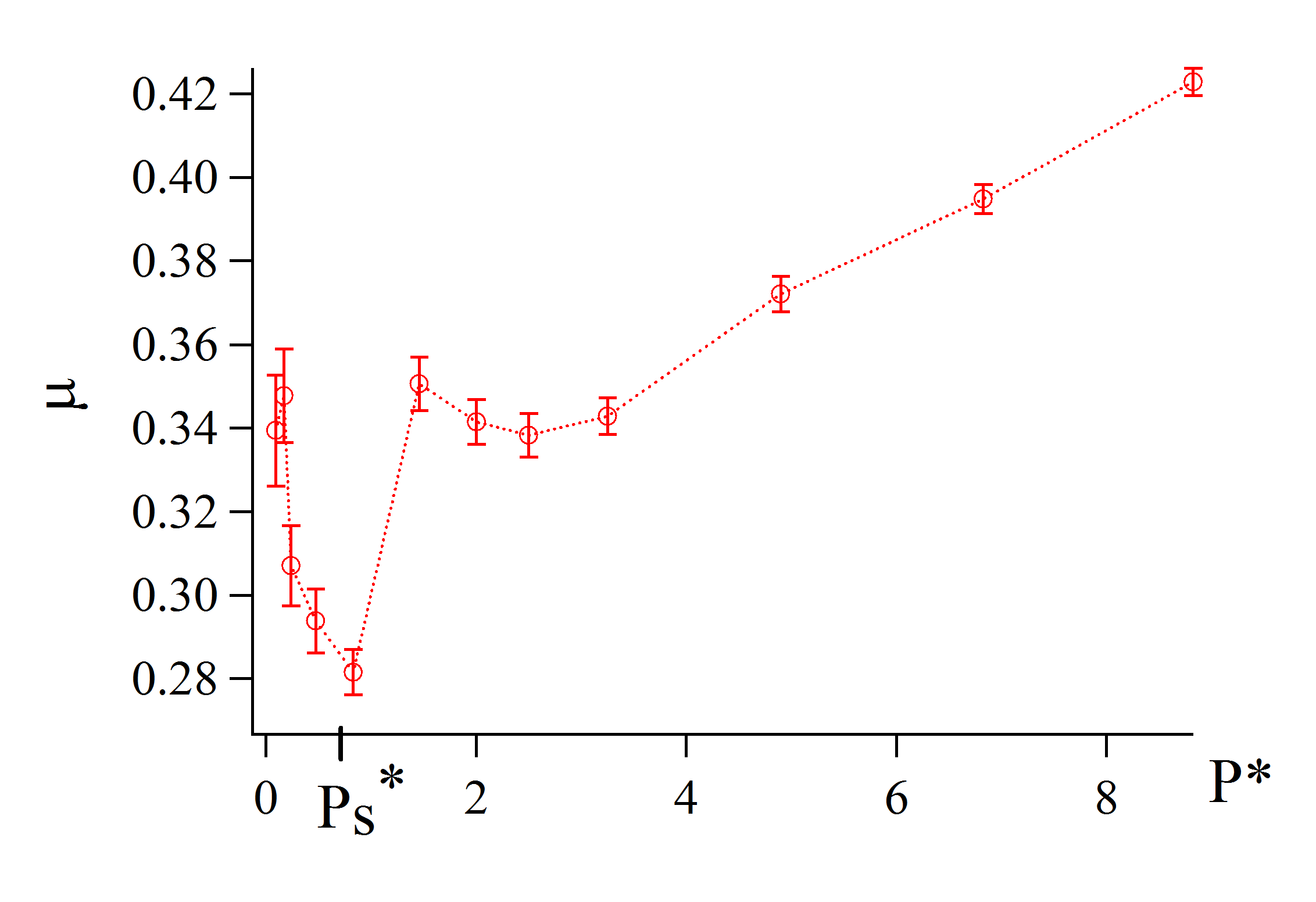}
\caption{Friction coefficient as a function of normalized applied pressure.}
\label{fig3}
\end{figure}

\begin{figure*}[!th]
\includegraphics[width=.45\textwidth]{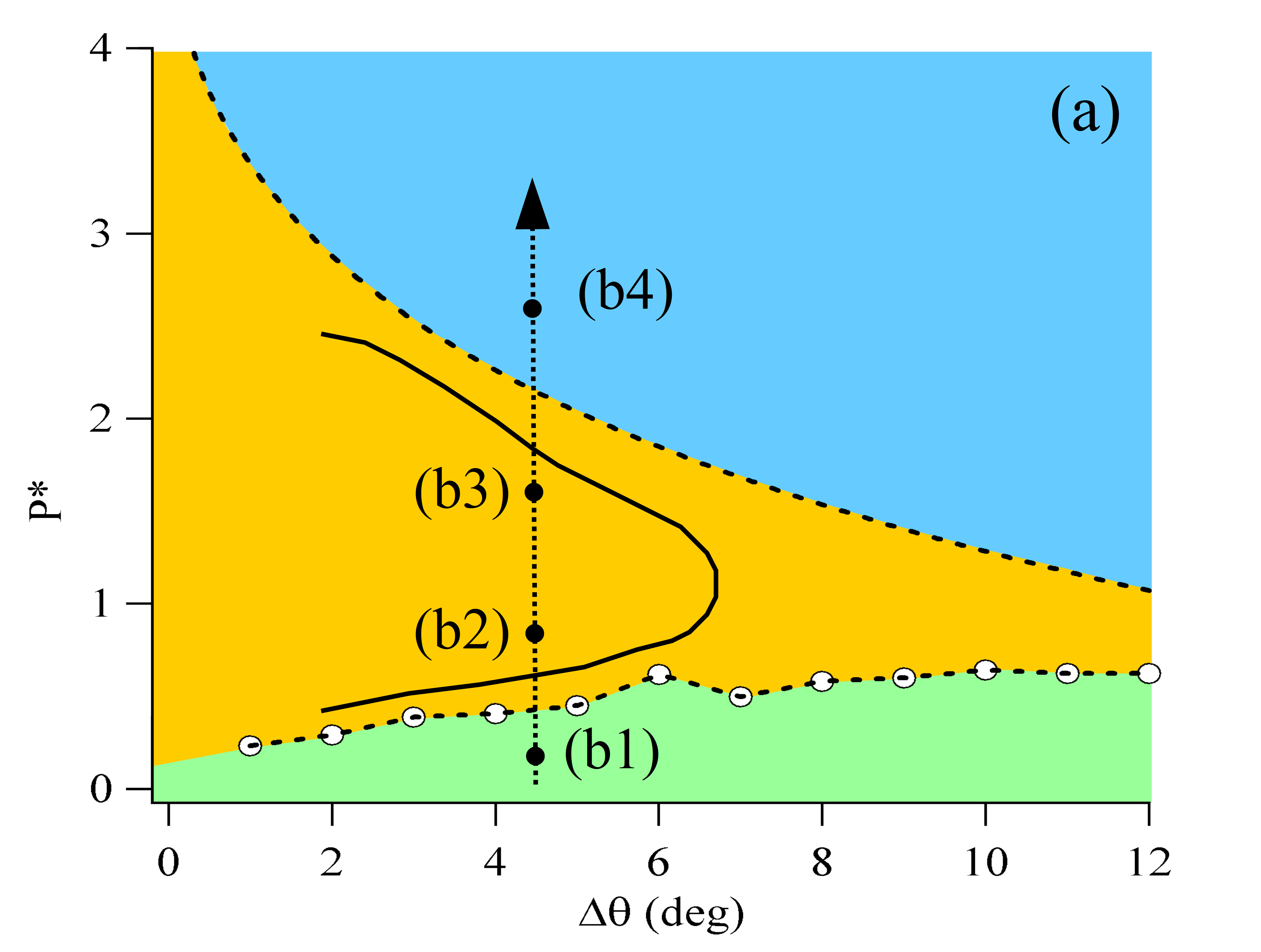}
\raisebox {8mm}{\includegraphics[width=.25\textwidth]{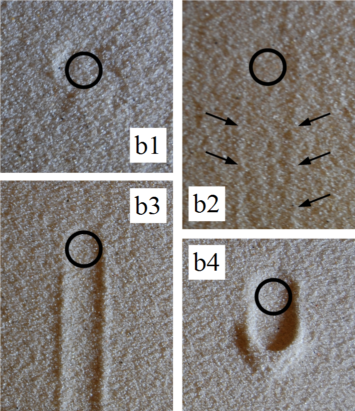}}%
\hspace{4mm}
\raisebox {20mm}{\includegraphics[width=.25\textwidth]{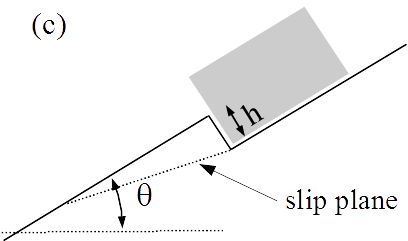}}
\caption{(a) Stability diagram for an object deposited on an inclined granular surface. The solid curve is the experimental sliding probability $\varphi=0.4$ for $\Sigma^*=30$. Dotted curve with circles is the lower stability threshold determined from DEM simulations. Dashed curve is the upper limit of stability deduced from the model of rim (see text). The points (b1) to (b4) stand for the pictures shown in (b). The colored domains represent the low $P^*$ (green) and high $P^*$ (blue) stability zones, and the sliding zone (orange). (b) Pictures of the tracks of deposited objects of different masses after removal or sliding. The circles represent the initial positions of objects (diameter $11~mm$), and number $1-4$ correspond to different zones of the stability diagram. In b1 the object was posed and removed without deformation, shallow and deep linear tracks are visible in b2 (arrows indicate the lateral rim) and b3. In b4, the object is stopped by a front rim of granular material. For reasons of contrast of photography, the material is natural sand ($d=300-400\mu m$) (c) Schematic view of an object letting a track under it. Dotted line represent the slip plane.}
\label{fig5*}
\end{figure*}

The dependence of the stability diagram with pressure suggests that the ratio between the tangential and the normal forces varies with the pressure applied to the granular material. This may be directly evidenced by measuring the friction coefficient $\mu$ for a slider ($\Sigma^*=35$) on a horizontal granular material ($d=2.2~mm$). For this, a slider is displaced at an imposed velocity $0.5~mm.s^{-1}$. We checked that decreasing further the velocity does not change the results. The force is measured with a force sensor and averaged on a total sliding distance of $600~mm$. Figure 4 shows $\mu$ for various applied pressure $P^*$. We observe a minimum value of $\mu$ around $P^*_s \simeq 0.6$. This is in agreement with the occurrence of a minimum of stability on inclined granular slopes. Previously published studies about friction on granular materials have been performed at larger values of $P^*\sim 10-100$ and did not report such variations of friction coefficients~\cite{geminard1999,fall2014}.

For understanding the physics governing the stability diagram, we observe the deformation of the surface for objects exerting different pressures. The solid curve in fig.5(a) represents the stability diagram. The dotted arrow represents a path of increasing $P^*$, and we shown in fig.5(b) pictures of the deformed surfaces along this path. For an object and a slope located at position (b1), the figure 5(b1), shows the lack of deformation created by an object deposited and then removed. In position (b2) the object slides, and we distinguish a shallow barely visible track (see arrows on fig.5(b2)).In position (b3) the object slides with a deeper, more visible track (fig.5(b3)). For heavier objects (b4), the deformation is so important that a rim of granular material forms in front of the object which stops its sliding (fig.5(b4)). So, the unstable zone of the instability diagram appears bounded for two different reasons.

We first discuss the transition between (b1) and (b2) which correspond to the onset of sliding. The extra load of the slider destabilizes slightly the granular slope. To our knowledge, the limit of stability of granular slopes in response to some localized force has not been reported in the literature. To investigate this point, we performed standard 2D Discrete Element Methods simulations. $4000$ particles (polydispersity $r_{max}/r_{min}=1.2$) are placed into a box which is tilted up to an angle $\theta<\theta_a=26~deg$. Disks interact via traditional linear, damped springs in normal and tangential directions, with a microscopic friction coefficient. An increasing localized vertical force is then applied to $6$ particles  until a destabilization (defined as a variation of the slope of $.05~deg$) of the surface occurs. The threshold {\jc is calculated for different $\Delta \theta$ by averaging over $20$ different packings}. The dotted curve with open circles of fig.5(a) shows the non-dimensional pressure necessary to destabilize the material as a function of $\Delta \theta$. This curve separates packings which are stable or unstable with respect to some localized pressure. {\jc The instability of granular slopes under an applied localized pressure is most likely to be understood as a special case of the more general problem of the stability of packing of frictional grains~\cite{vanhecke2007}.}

The transition between (b3) and (b4) corresponds to the formation of a track and of a rim in front of the slider which are large enough to stop the object. Figure 5(c) shows a schematic view of a slider which created a track. The motion of granular material in front of the slider creates a resistive force on the slider. This drag force may be calculated using the Coulomb method of wedges described in~\cite{nedderman}, the force being calculated here for an inclined granular material. The drag force $F_d$ exerted on a object of transverse length $L$ immersed of a depth $h$ into a granular material is then $F_d=(\rho g h^2 L / 2) \cos\theta f(\mu,\xi)$, where $f(\mu,\xi)$ is a function of the two parameters $\mu=\tan(\theta_a)$ and $\xi=\tan(\theta)$: $f(\mu,\xi)=(1+\mu \xi)/(1-\mu^2-4\mu t_m)$ with $t_m=[\sqrt{2 \mu (\mu-\xi)(1+\mu^2)(1-\mu \xi)}-2 \mu
(\mu-\xi)]/2 \mu (1+\mu \xi)$. For a horizontal material $f(\mu,0)=1/(\sqrt{\mu^2+1}-\sqrt{2}\mu)^2$~\cite{nedderman}, whereas close to avalanche angle $f(\mu,\mu)=(1+\mu^2)/(1-\mu^2)$. For a square object of surface $L^2$, the tangential component of the weight is $T=P L^2 sin(\theta)$. If $F_d>T$ the drag force is sufficient to stop the object. We found that the depth of the track is typically $0.4 P /\rho g$. This depth lie between the penetration depth of an intruder into a static granular material composed of glass beads: $\simeq 0.12 P / \rho g$~\cite{brzinski2013}, and the penetration depth into an Archimedean material of density $\phi \rho$, with $\phi=0.6$: $\simeq 1.7 P /\rho g$. There is a rim in front of the slider, whose height is of order of the track depth. So, in the following we will consider that $h=\beta P/\rho g$, with $\beta=0.8$. The condition of arrest for the slider $F_d>T$ may then be written as: $P^*> 2 L \cos \theta \sin \theta / d \beta^2 f(\mu,\xi)$. The dashed line of Fig.5(a) shows this stop condition calculated for $L/d=6$.

We now discuss the application of this stability diagram to animal locomotion. Indeed, the mechanisms identified on inert objects should apply to dynamical objects, even if additional phenomena specific to living animals of highly complex geometry and interactions with the ground will undoubtedly make the picture more complex. Close to avalanche angle, the stability of animals should depends on the animal's weight. By construction, the slope angles of the antlion trap is close to the avalanche angle~\cite{Griffiths1980a,Fertin2006}. So, we expect that a prey entering into the trap experiences a stability which depends on the pressure that it exerts.

 \begin{figure}
\includegraphics[width=.8\columnwidth]{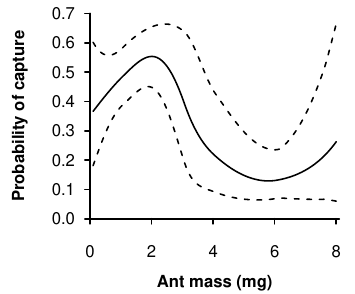}
\caption{The probability of ant's capture as a function of the ant mass (plain line) and it's confidence interval (dotted lines). Adapted from~\cite{Humeau.etal2015}.}
\label{efficiency}
\end{figure}

This mechanism agrees to the predation of ants by the antlion \textit{Euroleon nostras} (Geoffroy in Fourcroy, 1785) build traps into a granular material made of glass beads ($d=250 ~\mu m$)
(see fig.1(b)). Indeed, the efficiency of the trap has been estimated by the probability of capture of a prey that fall one time in the trap for different ant species having different weights~\cite{Humeau.etal2015}. Figure 6 shows the probability of capture as the function of the ant's mass. A maximum of capture occurs around $2~mg$ (For ant's mass $>7~mg$ the number of data points are small, as observed by the widening of the confidence intervals.). Neither species studied in~\cite{Humeau.etal2015} display morphological adaptation for walking on sand~\cite{Carothers1986,Higham2015,Seifert2017}, and we assume that the ants are scale-invariant. Noting $l_{ant}$ the ant's length, we should have $m_{ant}\sim l_{ant}^3$ and the surface of contact $\Sigma \sim l_{ant}^2$, and so $P^*\sim m_{ant}^{1/3}$. The occurrence of a maximum capture for an intermediate mass is then in agreement with the minimum of stability for an intermediate pressure $P^*_s$ that we evidenced. {\jc The recovery of stability at large applied pressure can be seen on fig.1(b), where a firebug does create footprints, as humans do. Those qualitative observations are in agreement with separated measurements of ant's~\cite{humeau.to.be.published} and firebug walks on inclined sandy surfaces.}

As a conclusion, we demonstrated that the stability of an object on an inclined granular surface is a subtle problem. The sliding is indeed restricted to a narrow range of applied pressure and angle. The instability occurs only when the surface may be slightly deformed by the slider weight, but not enough to create a rim able to stop the object. {\jc Our results are therefore in contradiction with Amontons-Coulomb laws, which need to be revisited to address friction phenomena on granular material at small applied pressure.} The existence of this instability diagram is expected to be relevant in many situations such as in robotic locomotion on sandy surfaces or in the formation of natural structures.

The work done by A.H. was part of his Ph.D. under the supervision of J.Ca. We thank the Région Centre and the University of Tours for financial support. J.Cr. thanks Sean Mc Namara \& Laurent Courbin for critical reading of the manuscript.

\bibliography{Biblio}{}
\bibliographystyle{unsrt}
\end{document}